# Saving Storage Space Using Files on the Web

Kevin Saric
QUT, CSIRO Data61
Brisbane, Australia
kevin.saric@hdr.qut.edu.au

Gowri Sankar Ramachandran
QUT
Brisbane, Australia
g.ramachandran@qut.edu.au

Raja Jurdak
QUT
Brisbane, Australia
r.jurdak@qut.edu.au

Surya Nepal
CSIRO Data61
Sydney, Australia
surya.nepal@data61.csiro.au

## ABSTRACT

As conventional storage density reaches its physical limits, the cost of a gigabyte of storage is no longer plummeting, but rather has remained mostly flat for the past decade. Meanwhile, file sizes continue to grow, leading to ever fuller drives. When a user's storage is full, they must disrupt their workflow to laboriously find large files that are good candidates for deletion. Separately, the web acts as a distributed storage network, providing free access to petabytes of redundant files across 200 million websites. An automated method of restoring files from the web would enable more efficient storage management, since files readily recoverable from the web would make good candidates for removal. Despite this, there are no prescribed methods for automatically detecting these files and ensuring their easy recoverability from the web, as little is known about either the biggest files of users or their origins on the web. This study thus seeks to determine what files consume the most space in users' storage, and from this, to propose an automated method to select candidate files for removal. Our investigations show 989 MB of storage per user can be saved by inspecting preexisting metadata of their 25 largest files alone, with file recovery from the web 3 months later. This demonstrates the feasibility of applying such a method in a climate of increasingly scarce local storage resources.

## 1 INTRODUCTION

Conventional computer hardware cannot improve forever. As current manufacturing processes hit their physical limits, the cost to store a gigabyte of data is no longer plummeting, but rather, has remained mostly constant for over a decade [11]. This is reflected in computer hardware configurations. The Macbook Air, likely the world's best-selling laptop, has been sold with a 256 GB storage option from 2010 to present (2024) [3, 4].

Magnetic platter Hard Disk Drives (HDDs) have traditionally improved due to areal density scaling, yet as the area required to store a bit becomes small enough, it begins to be affected by superparamagnetic forces, where the integrity of the bit fails [12]. These effects are reflected in the flattening price of small HDD storage, as seen in Figure 1. Similarly, as the transistors so fundamental to Solid-State Drives (SSDs) and flash drives become sufficiently small, they exhibit quantum tunneling effects, rendering them unreliable [18]. Moore's law, which in 1965 predicted the growth of the number of transistors in an unit of area [16], was predicted by Gordon Moore himself to come to an end around 2022 [18].

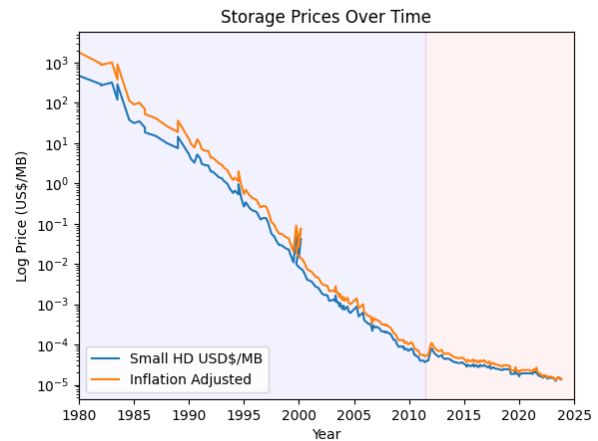

Figure 1: Log scale hard drive prices 1980s to present — note the flattening of storage prices 2012 to present

Cloud storage has been touted as the solution to constrained storage resources, in a move back to centralized control of data. Yet, despite the efficiencies offered by sophisticated cloud storage technologies, users face a loss of power to control storage costs as they become progressively more dependent on their particular vendor [17]. There is also the often-overlooked risk of entrusting files to a company that may someday collapse, or choose to put profits before privacy.

Replacement local storage technologies like various Storage Class Memory (SCM) solutions and DNA-based storage are still firmly in the research stage, and the question of which technology will supersede contemporary storage technologies remains unanswered.

In the midst of this evolving landscape state of data storage, user file sizes continue to grow [10, 15, 20]. These factors increasingly lead to the common scenario of a full drive for users. In these instances, the user must disrupt their workflow to laboriously remove old files to free space so their device can continue to function. This necessity is amplified when the user responsibly conducts backups, as the storage bloat propagates to backup drives. Thus, in absence of a breakthrough storage technology, users would benefit from more efficient, automated management of their largest files.

One way to reduce the storage burden of large files is to identify when they are available for recovery elsewhere. In these cases, the user can make a choice to keep a full redundant copy of the large file or to recover it from other sources. Filesystems and browsers typically – and without users' knowledge – record metadata about the origin source of downloaded files. On Windows systems this metadata is stored in Alternative Data Streams (ADSs) and on Unix-like systems, Extended Attributes (xattrs). A user, motivated to free up space but disinclined to do so laboriously, could authorize the deletion of an automatically-selected set of files which have been determined to still be available at their original sources on the web, as recorded in the ADS or xattr metadata. In doing so, the users accept the risk that those files may disappear from the web in the future — although this risk is mitigated by archival websites like the Internet Archive [2] and, often, mirror download sites.

In this paper we investigate users' filesystems for their biggest files and determine if those files happen to be available on the web. From these findings we propose a method of extending the paradigm of *recipes* — that is, obtaining data ingredients, combining them according to instructions, and testing the final result — to automate the freeing of space on a user's drive. We then evaluate the efficacy of our method.

The contributions of this paper are thus:

(1) A study and analysis of the 25 biggest files on a sampling of 10 participants' computers;
(2) A proposed method for conducting user-authorized file purging when files are available on the web; and
(3) An evaluation of our proposed method using 180 real-world local and web files.

Section 2 investigates the largest files on users' computers. Section 3 presents the model for user-authorized, automated file deletion. Section 4 evaluates the method. Section 5 discusses observations arising during the course of research. Section 6 presents related literature. Finally, Section 7 concludes the work.

## 2 BIGGEST LOCAL FILE ANALYSIS

In order to conduct this research, a dataset containing information about the biggest files on users' computer was required. Extensive effort went into locating a preexisting dataset to serve this need but none were suitable. Several were identified, including [9], [1], and [6]. Two of these were excluded for not collecting ADS or xattr origin source metadata. The final dataset custodian did not reply to emails.

This research thus set out to gather data directly from computer users via the Prolific study platform which allows researchers to screen and recruit study participants via their web site [5].

### 2.1 Selection Methodology

Participants were Windows users of adult age who were requested to run a PowerShell script on their computer which gathers the largest 25 files in storage, along with a list of the free and used space on their computers' drives. The 10 recruited participants were instructed to copy and paste the output of the PowerShell script into a survey form, which also contained a question asking their profession.

In order to ensure homogeneous results, only Windows users were recruited to participate. Children were excluded from the study. The number of largest files (25) was chosen through prior empirical testing to avoid overtaxing the participants' computers when running the script.

Ethics/IRB approval was granted according to the terms of the study. Participants were informed their participation is voluntary and encouraged to assess the output of the PowerShell script for private information before submitting. Only 1 participant chose not to provide information and exit the study and thus is not included, leaving 9 participants who provided results.

### 2.2 Script Function

The function of the PowerShell script provided to participants is described in commented code of Algorithm 1. The script first outputs basic information on the participants' storage drives (Line 3), then proceeds to find their 25 biggest files (Line 4). For each of these files, the name and ADS metadata are outputted (Lines 7 & 8).

**Algorithm 1:** PowerShell script provided to participants

```
   clear ;                              # clear PowerShell window
 1 Write-Host "STARTING" ;   # signal to user script is beginning
 2 cd / ;                           # change to root directory
 3 Get-PSDrive -PSProvider FileSystem ; # print storage drive info
 4 Get-ChildItem -Path $env:USERPROFILE -Recurse -File
     -ErrorAction SilentlyContinue |    # pass all user files to ...
     Sort-Object -Property Length -Descending | # ... be sorted ...
     Select-Object -First 25 |      # ... selecting the top 25 largest
 5 ForEach-Object {              # for each of top 25 largest files
 6     $file = $_.FullName ;                         # get file
 7     ls $file ;                              # print file name
 8     Get-Content -Path $file -Stream Zone.Identifier
         -ErrorAction SilentlyContinue ;     # print ADS metadata
 9 } ;
10 Write-Host "DONE" ;       # signal to user script is finished
```

### 2.3 Data Gathered

For each of the participants, a unique, anonymized ID was automatically generated by Prolific. For each of these, the top 25 largest files were identified and the following was collected for each file:

- Full names (including extension)
- Full path
- Size in bytes
- Last modified date
- ADS metadata, including:
  - `Zone.Identifier` (ZoneId) (where available)
  - `ReferrerUrl` (RU) (where available)
  - `HostUrl` (HU) (where available)

ZoneId, RU and HU are fields within the ADS metadata that are often — and optionally — populated by software with information about the source of the file, usually Uniform Resource Locators (URLs). Owing to the voluntary use of these fields, they are at times not used.

Saving Storage Space Using Files on the Web

## 2.4 Study Results

**Table 1: Biggest file analysis summary table**

| Parameter | min | mean | median | max |
|---|---|---|---|---|
| Participants (P) | | | 9 | |
| Total files | | | 180 | |
| Files per P | 1 | 18.1 | 25 | 25 |
| File size | 4 MB | 796 MB | 163 MB | 27 GB |
| Days since last modified | 3 | 555 | 579 | 1595 |
| File name length (ex-extension) | 3 | 27 | 24 | 92 |
| Inter-P duplicated files | | | 0 | |
| Intra-P duplicated files | | | 6 | |
| ZoneId reported | | | 79 (51.0%) | |
| Only ReferrerUrl (RU) | | | 5 (3.2%) | |
| Only HostUrl (HU) | | | 42 (27.1%) | |
| Both RU and HU | | | 30 (19.4%) | |
| Neither RU nor HU | | | 78 (50.3%) | |

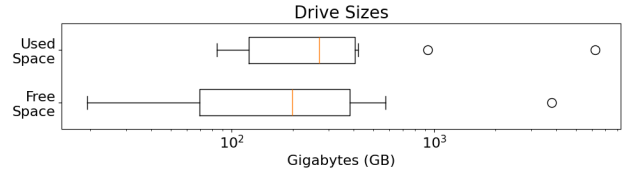

**Figure 4: Participant's drives sizes and utilization**

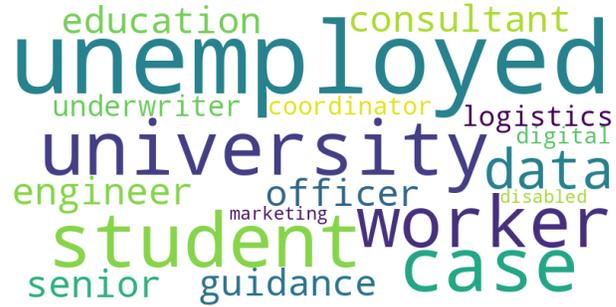

**Figure 5: Reported professions of participants**

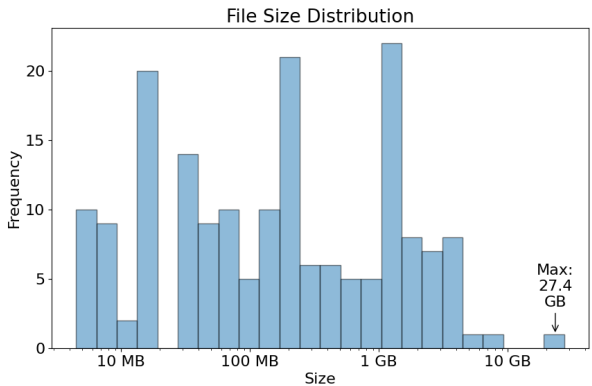

**Figure 2: File size distribution**

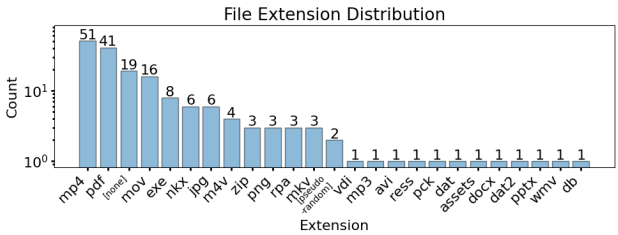

**Figure 3: Extensions of reported files**

The results of the study are listed in Table 1 and Figures 2 to 5. Other studies have provided general statistical observations of users' files; we have provided these too for completeness. Yet, in the context of saving space using files available on the web, our focus is on investigating the usage of RU and HU fields in real-world files so as to understand the extent they can be relied upon to recover files from the web. The observed usage of these fields is reported below, categorized logically. Their groupings arise from inspecting the participants' supplied script outputs and manually interpreting their RU and HU records. Typically this involved manually accessing the URLs and investigating their content and/or inferring the underlying content from the composition of the RU/HU URL.

Categorizing in this way provides an indication of the scale of space can be saved using file recipes per category. Some RU and HU URLs may point to files widely available on the web for anyone to obtain. Others may link to files available from Cloud Storage Providers (CSPs), for example, implying the authorized user can access the file at a later date, but others user cannot due to login requirements. These factors are mentioned here for context and are further explored in the Evaluation section — particularly Table 3 and Table 4 — where we attempt to reconstitute the participants' files from their reported RU and HU records. Here we report primarily on the population of the RU and HU metadata fields which has not previously investigated in academic research to our knowledge.

In total, 180 files were reported. Of the 9 participants that provided results, 7 responded with all 25 of their largest files, while 2 participants only supplied 4 and 1 of their biggest files respectively due to a script execution error — likely due to insufficient user privileges. In all instances where a ZoneId was reported (51% of files) it was listed as Zone 3, referring to the Internet. In all other cases, no ZoneId was reported.

*2.4.1 Cloud Collaboration.* Microsoft services were commonly seen in the results. According to both RU and HU records, 3 files were from SharePoint URLs across 2 participants and 3 other files were sourced from Teams domains listed in the RUs across 2 participants, including 1 instance of a Teams installer.

The frequent discovery of cloud collaboration platform URLs in the RU and HU fields suggests that some files will only be available only to the file's owner via a login or authentication process. Furthermore, if the user is not the file owner, their access may be revoked in the future. This observation necessitates the division of results in the Evaluation section of this paper into *Publicly Redownloadable (Public Rd)* and *Redownloadable with Authentication (Rd w Auth)* sections, as will be evident later.

*2.4.2 Webmail.* A single participant sourced 8 files from `https://outlook.office.com/` as listed in both their RU and HU records. This presumes the need for the user to remain authorized to access the account and not have deleted the file, similar to that discussed in Section 2.4.1, but via a distinctly different access mode, webmail.

*2.4.3 Big Tech CSPs.* Despite focusing on Windows, 6 iCloud files were reported in the RUs for a single participant, with a full path to the source in the HU. This also presumes the need for the user to remain authorized to access the account and not have deleted the file, similar to that discussed in Section 2.4.1 and Section 2.4.2, but via a distinctly different access mode, cloud storage.

*2.4.4 Small CSPs.* A university cloud storage platform accounted for 4 files with 1 participant according to their HU records. Less well known file sharing sites — ones not run by "Big Tech" firms such as Apple, Google and Microsoft — contributed 2 files across 2 participants, also as evidenced by their HU records. Unlike the CSPs in Section 2.4.3, small CSPs tend to have a wider range of access mechanisms, including, at times, direct public accessibility.

*2.4.5 Applications/Tools.* A single participant listed 25 files from a video downloader website URL according to the HU records. Meanwhile, a single participant listed 15 files with a PDF tool website URL in the RU. Similarly, another single participant reported 3 files that began with `chrome-extension://` in the HU only. These types of tools limit visibility into the underlying file by using these fields as a sort of promotional channel, despite the fact they are seldom seen by users.

*2.4.6 Local Access.* With one participant, 5 files had RU records that began with `C:\` thus being reported from a local file, 3 of which were `.rpa` files — common in graphic novels — and 1-2 GB in size. Of these 3, all were listed as being originally sourced from local `.zip` files according to the RUs. Additionally, 5 files from a single participant began with `file://` in the HU, but without any record in the RU. In all these 5 cases, the HU terminated in a common video file format extension and were located in a path containing the string `/Downloads/`. This suggests considerable storage usage by some users is attributable to downloaded — and presumably redownloadable — videos rather than self-generated content.

*2.4.7 Direct Links.* With direct links, the URL to redownload the file was immediately available via the associated HU record. Of the 11 files with direct web links, 6 were also accessible via the RU, however this required loading the page in-browser and finding the link already provided more directly in the HU. These links are particularly amenable to file recovery from the web, given they are easily accessible to any user with web access.

*2.4.8 Links Not Recorded.* Of the 180 participants' files, 88 were listed with neither a RU nor a HU record. Therefore, none of these can be restored from source using only their ADS origin source metadata.

Across all the categories, the largest files were primarily video files, including several movies, presumably ripped or downloaded. Of these, all had blank RU and HU records, suggesting the possibility of them being acquired through P2P or torrent services. Nevertheless, one user had 6 files with the string "Zoom meeting" included in the file name. Video files were broadly popular with 6 participants holding them.

Of the 180 total reported files, 8 were executable programs (including installers) held by 3 participants. Of these, 6 listed neither RU nor HU records. There were also 3 `.zip` files across 2 participants. Of these files, 2 listed a HU record only. This file type suggests the production of multiple smaller files across the participants' their drives, provided they were extracted.

The findings presented in this section suggest several design considerations for saving storage space. The prevalence of a small number of large files, mostly videos, suggests that resources should primarily be allocated to evaluating the largest files for availability on the web. Additionally, since the application of the HU and RU records is inconsistent, both should be evaluated for file redownloadability, especially since both records are very lightweight to query. In the next section these observations are applied.

## 3 WEB-BACKED FILE PURGING METHOD

The insights derived from the study in Section 2 enabled the next step: crafting a means of freeing space that leverages file availability on the web as a recovery source.

### 3.1 Recipes

*Recipes* have metaphorically existed in computer science for so long as to not have a definite origin. Yet, in the context of storage, it the terms is widely reported to have originated with Tolia et al. at Intel Research in 2003 [24].

Conceptually, recipes imply obtaining data components and processing them according to some given instructions. The term is widely used in the field of deduplication, which itself is closely related to this work. In deduplication, redundant copies of the same data are often replaced with a smaller recipe for reconstituting the data from other sources at a later date. Appropriately, this makes the recipes metaphor fitting in the context discussed here.

Adapting recipes to reference web-sourced data requires an understanding of the web and its behaviors. We do not make predictions of the availability of files in this work. Such predictions have been explored in other work [21] and are out of scope of this work. Here we focus on recovering files from their original sources on the web as reported by the file's ADS origin source metadata records.

### 3.2 Model

Here we propose the following model:

- A local file $F$ on the user's computer/device;
- Recipe $R$, derived from $F$ and its ADS/xattr metadata;

- Unobstructed access to a duplicate of $F$ on the web, represented as $F_w$.

We define three stages: (i) *creation* — where $F$ is represented as $R$ for the first time; (ii) *maintenance* — where $F_w$ is confirmed to remain available on the web; and, (iii) *reconstitution* — where $F$ is generated from $F_w$ using $R$. When $F$ is reconstituted from web sources, it is denoted as $F_r$.

Although it's widely known that "chunking" a file increases the benefit of deduplication, it has been shown throughput is reduced by a factor of 15 [7] in order to achieve a merely modest increase in space savings. For that reason this model focuses on whole files rather than chunks, knowing that often that whole file is archived. This makes the file the atomic unit of this work.

## 3.3 Data Security

Here we address the hallmarks of data security: Confidentiality, Integrity & Availability (CIA).

*3.3.1 Confidentiality.* During the recipe creation phase, confidentiality is assumed through End-to-End Encryption (E2EE) where the recipe is encrypted on the user's computer to prevent eavesdropping and tampering and ensure confidentiality when copying/moving $R$ between systems. This is denoted as $\text{e2ee}(R)$. Recipes are thus exchanged with semantic security, where the transmission of the recipe cannot be inferred by an intercepting party. During the reconstitution phase, confidentially may be impacted by observing the download of the file from the web using a Man-In-The-Middle (MITM) attack. This is acceptable on the assumption that users will be willing to download a file that they likely downloaded in the first place. Should they require additional confidentiality, mixnets, proxies or Virtual Private Networks (VPNs) can be employed, thus gaining confidentiality against MITM snooping during reconstitution in exchange for performance.

*3.3.2 Integrity.* Integrity is ensured through comparing the cryptographic hash (i.e. identifying fingerprint) of the original file $F$ with that of the reconstituted $F_r$. During the creation phase, this necessitates hashing $F$ using a suitably strong hashing algorithm, and storing the hash in recipe $R$. Upon reconstitution, the original hash is obtained from $R$, and if $\text{hash}(F) == \text{hash}(F_w)$, file integrity is assumed.

*3.3.3 Availability.* Often, filesystems and software interact to track the original source of downloaded files. In Windows systems, this is reported in the ADS metadata while on Unix-like systems, this is stored as an xattr metadata record, denoted $\text{src}(F)$ in either case. This provides the source of greatest likelihood of availability for reconstitution. As such, this information is included in $R$ upon creation. Upon reconstitution, the original source may have changed. This may, in some instances, necessitate finding alternative sources $F$. This function can be provided by external services [21, 22] and is beyond the scope of this paper.

## 3.4 Recipes Content

We propose an approach to recipe creation that includes the following information, encapsulated in a data format such as JSON:

- Date/timestamp of recipe creation

Table 2: How Confidentiality (C), Integrity (I) and Availability (A) are addressed by Files ($F$) & Recipes ($R$) in the system model

| Factor | Phase | Comments |
| --- | --- | --- |
| C | $R$ creation | $\text{e2ee}(R)$<br>Recipe is encrypted client-side |
| C | $R$ maintenance | Recipe remains encrypted |
| C | $F$ reconstitution | Use mixnet/VPN if required |
| I | $R$ creation | $R \leftarrow \text{hash}(F)$<br>Hash of file stored in recipe |
| I | $R$ maintenance | Recipe storage integrity assumed |
| I | $F$ reconstitution | Confirm $\text{hash}(F) == \text{hash}(F_w)$<br>File is identical to original |
| A | $R$ creation | $R \leftarrow \text{src}(F)$<br>Source of file stored in recipe |
| A | $R$ maintenance | Local recipe availability assumed |
| A | $F$ reconstitution | $\text{src}(F)$<br>Request file from original source |

- Date/timestamp of latest recipe maintenance
- ADS/xattr source URLs (any available, e.g. RU and HU)
- Original file path
- File name including extension
- File size in bytes
- Full cryptographic hash of file
- Cryptographic hash algorithm used (e.g. SHA256)

Additional information can be included to offer enhanced metrics or to optimize performance. One example by be a partial hash (e.g. a hash of the first megabyte of a file) to support a fail-fast mechanism for verifying the integrity of a reconstituted file. These enhancements are encouraged but not discussed fully here for brevity and due to their obviousness.

## 3.5 Proposed Processes

*3.5.1 Creation.* Figure 6 outlines the proposed steps used to free up storage space by creating small recipes to replace large files using this model. These steps are run in software which provides these functions. They are as follows:

Upon starting the software, the 1$^{\text{st}}$ largest file on disk is found (i). The $\text{src}(F)$ (both RU and HU) is obtained from the ADS/xattr metadata (ii). The $\text{src}(F)$ is then requested from the server (iii). If the server responds with the file, it is compared against the local file to ensure it is an identical match (iv). If a hash of the file provided by the server matches a hash of the local file, and the user authorizes it, then a recipe is created in place of the file, thus purging the large file (v). This process is continued until sufficient free space is created (vi). An alternative case exists where the file is not directly downloadable, but can be found by examining the links on a ADS/xattr-provided web page. In this instance (vii) the page is scraped for indirect links which themselves may provide a copy of the original file.

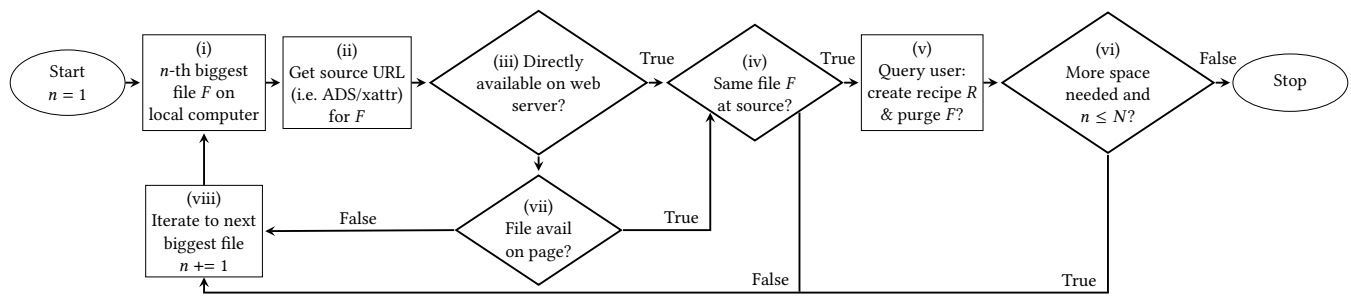

Figure 6: Process for using files available on the web to free storage space

*3.5.2 Maintenance.* On a regular schedule, recipes should be evaluated for currency to inform the user if their files are no longer available at the addresses originally found in the ADS/xattr metadata. In these circumstances, the user may be provoked to manually find these files at alternative source. This may be done manually or using a service designed for this purpose [21, 22].

*3.5.3 Reconstitution.* When attempting to reconstitute files from the original ADS/xattr sources on the web, the process is similar to steps (iii), (iv) and (vii) in Figure 6. Namely, the file is retrieved from the web source, hashed and the process is completed only if the cryptographic hash of the downloaded file matches that of the original local file.

## 4 EVALUATION

With the recipe content and processes (Sections 3.4 to 3.5) defined, we are able to evaluate the data provided in Biggest Local File Analysis, (Section 2) according to the Web-Backed File Purging Method (Section 3). We begin by investigating the magnitude of potential space savings. This was accomplished by using the information provided by participants to attempt a recovery of their files from their original sources on the web.

Copies of the participants' files were requested from the reported ADS metadata (specifically the RU and HU) web addresses. If access to the file on the web was obstructed, the link was investigated manually by interpreting the URL. In some cases the URLs pointed to a web page which was checked manually for indirect access to the file, such as requiring a user click a button or link to download. When URLs pointed to CSP, webmail, or cloud collaboration software, access to the file for the user was presumed and noted as such.

This experiment was run 95 days, or approximately 3 months, after the study. Results are shown on Tables 3 to 4. These tables focus on tabulating the redownloadability results of the underlying files from the web. This is signified by if they are *Not Redownloadable (Not Rd)*, usually due to a lack of a public URL from which to request the file; *Publicly Redownloadable (Public Rd)* to those with web access; or *Redownloadable with Authentication (Rd w Auth)* where we presume the file can be again accessed for the user, given continued authorization to access the file. The columns correspond with the categories discussed in Sections 2.4.1 to 2.4.8.

The results were obtained using the network access of a major research organization in a developed nation and did not appear to suffer from any network obstructions.

The findings of this evaluation show significant variation in the use of the RU and HU records by software. This is a result of the lack of standardization or enforcement on their usage. Broadly speaking, the RU tended to link to a base domain or web page, while the HU linked more directly to the web file itself. However, out of 10 cases where the HU record linked directly to the expected file, the same file was also recoverable in 6 cases via the RU by traversing links on the returned web page. This suggests the HU record is the better link to first check when reconstituting files — hence (iii) on Figure 6 — with a backup indirect route potentially available via the RU — hence (vii) on Figure 6. The same principle applies regardless of whether the link is Public Rd or Rd w Auth

Of the Public Rd files, all were redownloadable via the link in their HU record. In 91% of cases these files also had a RU record, but in only 82% of cases did the RU record contain an indirect download link to the required file. For the Rd w Auth files, a RU record was available in 30% of the cases and a HU record available in 31% of the cases. All this implies a degree of redundancy between the utility of the RU and HU fields. This is likely an artifact of different software's varying behavior when populating the ADS record.

In the cases of small CSPs, a single link provided indirect access to the file, requiring the user to click to download, but without any CAPTCHA or onerous obstruction. In another small CSP case, the HU merely listed the small CSP's root web address, preventing direct access to the file. In the remaining 4 cases of small CSP HU links, the small CSP had ceased operation causing a timeout when requesting the file.

In the case of HU links beginning with `chrome-extension://`, the underlying behavior of the extension cannot be fully interpreted, however there is a likelihood that in all cases the extension was a PDF reader.

The single HU instance of a browser app appeared to be an in-browser zipping utility. Similarly, PDF and video converters websites proved popular but did not enable subsequent redownloading via the RU or HU metadata.

According to the RU record, a per-user average of only 5.1 MB ± 15.2 MB ($\not< 0$) of storage was publicly accessible, however via authorized access, a more substantial 818.2 MB ± 1.64 GB ($\not< 0$) was accessible. Via the HU record, a per-user average of 170.5 MB ± 511.4 MB ($\not< 0$) was publicly accessible, with 372.3 MB ± 1.08 GB ($\not< 0$) accessible via authorized access.

Table 3: File redownloadability of 180 biggest files after 3 months using ReferrerUrl (RU) metadata — *Public Rd*: files redownloadable by anyone; *Rd w Auth*: redownloadable by authenticated users, such as via a login.

| RU Link Desc. | Total Count | Not Rd | Public Rd | Rd w. Auth. |
|---|---|---|---|---|
| **Cloud Collaboration** | 7 | 5 | – | 2 (presumed) |
| **Webmail** | 8 | – | – | 8 (presumed) |
| **Big Tech CSPs** | 6 | 6 | – | – |
| **Small CSPs** | 1 | 1 | – | – |
| **Applications/Tools** | | | | |
| PDF converter | 15 | 15 | – | – |
| Search engine link | 2 | 2 | – | – |
| **Local Access** | | | | |
| `C:\` location | 5 | – | – | 5 (presumed) |
| **Direct Links** | | | | |
| Webpage URL | 6 | – | 6 (via links) | – |
| **Links Not Recorded** | 130 | 130 | – | – |
| **Total Rd in MB/GB** | – | – | 45.6 MB | 7.36 GB |
| ↳ **Per Participant** ($\not< 0$) | – | – | 5.1 MB ± 15.2 MB | 818.2 MB ± 1.68 GB |

Table 4: File redownloadability of 180 biggest files after 3 months using HostUrl (HU) metadata — *Public Rd*: files redownloadable by anyone; *Rd w Auth*: redownloadable by authenticated users, such as via a login.

| HU Link Desc. | Total Count | Not Rd | Public Rd | Rd w. Auth. |
|---|---|---|---|---|
| **Cloud Collaboration** | 7 | – | – | 7 (presumed) |
| **Webmail** | 8 | – | – | 8 (presumed) |
| **Big Tech CSPs** | 6 | – | – | 6 (presumed) |
| **Small CSPs** | 6 | 5 | 1 (req'd click) | – |
| **Applications/Tools** | | | | |
| PDF converter site | 15 | 15 | – | – |
| Video converter site | 25 | 25 | – | – |
| Chrome extension | 3 | ? | ? | ? |
| In-browser app | 1 | ? | ? | ? |
| **Local Access** | | | | |
| `file://` location | 5 | – | – | 5 (presumed) |
| **Direct Links** | | | | |
| Download URL | 11 | – | 10 | 1 (presumed) |
| **Links Not Recorded** | 93 | 93 | – | – |
| **Total Rd in MB/GB** | – | – | 1.53 GB | 3.35 GB |
| ↳ **Per Participant** ($\not< 0$) | – | – | 170.5 MB ± 511.4 MB | 372.3 MB ± 1.08 GB |

## 5 DISCUSSION

Most typical users, and even many avid computer users, are unaware of the existence of ADS/xattr metadata which records the origin sources of files altogether. Perhaps as a result, this is the first study known to the authors investigating their real-world usage. As a result of these new findings, this study also opens opportunities for saving storage space using files available on the web.

Some general observations can be made on the basis of the findings. For example, a small number of users store unusually large files, namely video and multimedia files as evidenced by the large per-participant file redownloadability standard deviations in Table 3 and Table 4, suggesting some users have use patterns that are better suited to recovery from the web. These users are disproportionately benefited from persistent, remote access to these files which are presumably only accessed a small number of times. Inversely, a larger number of users have many small to medium sized files (e.g. PDFs) that may be obtained in the course of their work and easily recoverable from the web at a later date. This particular subgroup of users is not fully characterized by this study as it was limited to the 25 biggest files on users' system. Nevertheless, given the negligible overhead of checking the ADS/xattr metadata and the source URLs, a large quantity of small to medium sized files can easily be assessed for purging as easily as with a small number of large files. The reason this was not done in this study was to avoid overburdening the data collection script required to run on participants' computers, but this would not be an issue in production software.

The per-participant file redownloadability standard deviations reported in Table 3 and Table 4 are large compared to their respective means. This represents a limitation of this work and suggests that future research may benefit from larger sample sizes.

Sampling bias was mitigated somewhat by use of the Prolific platform to recruit a wide variety of participants. This, paradoxically also introduced a bias towards users typically operating computers during the day and available to contribute to studies. Although Windows remains the dominant operating system, the number of MacOS, Linux and mobile users are not insignificant, yet have not been studied here.

The time period for trialing reconstitution of files from the web was three months. This is a singular evaluation point and further longitudinal studies would be needed over time to assess the techniques laid out here across other time frames.

Access authorization plays a role in the redownloadability of files. Many of the users' files were reported to be from CSP, cloud collaboration and webmail sources, suggesting the owners of the files are able to recover them at later dates, but not necessarily guaranteeing it. Visibility into the redownloadability of these files is limited in this study.

While it is tempting to equate the reconstitution rate at three months to the Service-Level Agreements (SLAs) of CSPs, attempts to do so were abandoned as misleading. The approach to storage management discussed here is quite unlike conventional ones, despite some other similar work in automatically obtaining files from the web [21, 25] and thus efforts to equate these metrics to industry standard defy direct comparison.

From the study evaluation in Section 4, we observed that there is significant variation in how software utilizes the HU and RU fields of the ADS metadata. Since these fields are not standardized and their usage is not validated or enforced, this enables a variety of usage techniques employed by developers. This suggests that neither field can be independently relied upon to find the source of a file, and that while the HU field has greater reconstitution rates, URLs from both fields should be checked to locate a copy of the file.

Being an initial study, many opportunities for future work exist. These include considering a larger number of files, a greater number of participants and more system configurations (e.g. MacOS). Additionally, this study opens questions of how the web itself can better support the space savings needs of users in a communal way, such as by providing files at a later date.

## 6 RELATED WORK

Saving storage space can occur using a variety of techniques, most notably compression and deduplication. Compression finds redundant patterns internally — within the data — and replaces these redundancies with smaller instructions on how to reverse the compression later on [26]. Deduplication, by contrast, finds redundant patterns externally — across different files — and replaces those duplicated files with smaller instructions about how to reverse the deduplication later on [23].

Deduplication has grown to be one of the most essential storage-saving techniques used by CSPs [23]. With the users of a CSP uploading a wide variety of files, these files are often identical copies of other users' files. When these duplicate files are detected by the CSP, they will normally only store a single copy to save the expense of unnecessary storage hardware. This approach has been found to result in storage savings of greater than 72%, far superior to compression techniques in many CSP implementations [15].

While the elimination of redundant data has a long history in a variety of contexts [19], the conceptual description of "recipes" for deduplication is largely credited to Tolia et al. in association with Intel Research [24]. Tolia et al. developed the concept of recipes to support CASPER, a recipes-based file system. Over time, the conceptual use of recipes became commonplace and were increasingly modified to meet several different data management objectives. Notably, advancements made by Meister et al. advanced the field using a variety of deduplication recipe improvements and optimizations [14]. The concept of recipes is now so ingrained as to be mentioned throughout comprehensive surveys on the field of deduplication [26].

The application of these concepts to purging files according to their availability at their origin sources according to ADS/xattr records has not yet been undertaken in literature. While other research exists covering the identification of pages and documents on the web showing similarities [8, 13], these have largely been undertaken in the context of web search search engines and crawlers for the purposes of detecting changes to web page content.

Traeger et al. come closest to the concept described in this paper, describing a means to deposit files into web caches and webmail services with the intent of retrieving them later [25]. Yet this approach requires a proactive uploading of files rather than passively benefiting from these files being available on the web in advance. This paper exists to address this gap in prior research.

## 7 CONCLUSION

File systems track the origins of a file using preexisting metadata channels. This metadata can be used to determine the file's availability on the web. By doing so, users can make informed decisions about what files to purge when freeing storage space. We investigated the 25 largest files on Windows users' drives along with accompanying source metadata. Using this information we proposed a file purging method that leverages this metadata. Evaluating the results of our investigation and proposed method, we found that, on average, several hundred megabytes per user could be purged and recovered form the web in three months time. This work provides novel insights and proposes a means of enhancing backup and space-freeing software. It thus provides a means to counteract slowing storage advancements and growing storage constraints.

## ACKNOWLEDGMENTS

The work has been supported by the *Cyber Security Research Centre Limited* whose activities are partially funded by the Australian Government's *Cooperative Research Centres Programme*.